\documentclass{article}

\usepackage[affil-it]{authblk}
\usepackage[dvipsnames]{xcolor}
\usepackage{amsfonts}
\usepackage{amsmath,amsthm,amssymb,dsfont}

\usepackage{enumerate}
\usepackage{graphicx}	
\usepackage{subcaption}
\usepackage[margin=3cm]{geometry}
\usepackage{url}
\usepackage{todonotes}
\usepackage{bbm}

\usepackage{tikz}
\usepackage{circuitikz}

\usetikzlibrary{arrows.meta} 
\tikzset{ meter/.append style={ draw, inner sep=10, rectangle, font=\vphantom{A}, minimum width=30, line width=.5, path picture={ \draw[black] ([shift={(.1,.3)}]path picture bounding box.south west) to[bend left=50] ([shift={(-.1,.3)}]path picture bounding box.south east); \draw[black,-{Latex[scale=0.6]}] ([shift={(0,.1)}]path picture bounding box.south) -- ([shift={(.3,-.1)}]path picture bounding box.north); } } }

\usepackage{pifont}
\usepackage{multirow}
\usepackage{makecell}

\usepackage{epsfig}
\usetikzlibrary{shapes.symbols,patterns} 
\usepackage{pgfplots}
\pgfplotsset{compat=1.10}
\usepgfplotslibrary{fillbetween}
\usetikzlibrary {decorations.pathmorphing, decorations.pathreplacing, decorations.shapes}

\definecolor{linkblue}{HTML}{001487}
\usepackage{hyperref}
\hypersetup{colorlinks=true,citecolor=linkblue,linkcolor=linkblue,filecolor=linkblue,urlcolor=linkblue,breaklinks=true}

\usepackage{nicefrac}
\usepackage{mathtools}

\usepackage{thmtools} 
\hypersetup{hypertexnames=false}

\usepackage{algorithm}
\usepackage{algorithmic}

\usepackage{mdframed}
\usepackage{aligned-overset}
\usepackage{circuitikz}

\theoremstyle{plain}
\newtheorem{theorem}{Theorem}[section]
\newtheorem{lemma}[theorem]{Lemma}

\newtheorem{corollary}[theorem]{Corollary}

\theoremstyle{definition}

\newtheorem{remark}[theorem]{Remark}

\newtheorem{non-example}[theorem]{Non-example}

\DeclareRobustCommand{\abbrevcrefs}{%
\Crefname{theorem}{Thm.}{Thms.}%
\Crefname{corollary}{Cor.}{Cors.}%
\Crefname{lemma}{Lem.}{Lems.}%
\Crefname{remark}{Rmk.}{Rmks.}%
\Crefname{proposition}{Prop.}{Props.}%
\Crefname{equation}{Eq.}{Eqs.}%
\Crefname{example}{Ex.}{Exs.}%
}

\DeclareRobustCommand{\Cshref}[1]{{\abbrevcrefs\Cref{#1}}}

\newcommand*{\PP}{\mathbb{P}}
\newcommand*{\E}{\mathbb{E}}
\newcommand*{\ee}{\mathrm{e}}

\newcommand*{\cC}{\mathcal{C}}
\newcommand*{\cD}{\mathcal{D}}

\newcommand*{\cH}{\mathcal{H}}

\newcommand*{\cJ}{\mathcal{J}}

\newcommand*{\cN}{\mathcal{N}}
\newcommand*{\cM}{\mathcal{M}}
\newcommand*{\cP}{\mathcal{P}}

\newcommand*{\cR}{\mathcal{R}}
\newcommand*{\cS}{\mathcal{S}}

\newcommand*{\N}{\mathbb{N}}

\newcommand*{\R}{\mathbb{R}}

\newcommand*{\C}{\mathbb{C}}

\newcommand*{\LOCC}{\mathrm{LOCC}}

\newcommand*{\St}{\mathrm{S}}

\newcommand*{\eps}{\varepsilon}

\newcommand*{\id}{\mathds{1}}

\newcommand*{\supp}{\mathrm{supp}}
\newcommand*{\tr}{\mathrm{tr}}
\newcommand*{\ket}[1]{| #1 \rangle}
\newcommand*{\bra}[1]{\langle #1 |}

\newcommand{\proj}[1]{|#1\rangle\!\langle #1|}

\newcommand*{\di}{\mathrm{d}} 

\newcommand{\norm}[1]{\left\lVert#1\right\rVert}

\usepackage{float}
\usepackage[nameinlink,capitalize,noabbrev]{cleveref}

 \allowdisplaybreaks

\definecolor{protocolblue}{HTML}{1F5A94}
\definecolor{protocolorange}{HTML}{C66A2B}
\definecolor{protocolgreen}{HTML}{2F7D68}
\definecolor{protocolgray}{HTML}{68717A}

\title{Distillable entanglement is continuous} 

\author[1,2]{Lukas Schmitt}
\author[2]{David Sutter}
\affil[1]{\small Institute for Theoretical Physics, ETH Zurich}
\affil[2]{\small IBM Research Europe -- Zurich}
\date{\vspace{-1cm}}

\date{}

\begin{document}

\maketitle

\begin{abstract}
We prove continuity of distillable entanglement, the optimal asymptotic rate at which copies of a bipartite state can be converted into maximally entangled pairs by local operations and classical communication (LOCC) with vanishing error. We establish an explicit asymptotic continuity bound that holds for arbitrary bipartite states and depends logarithmically on the system dimension. This bound has the same structure as known continuity bounds for conditional entropy, entanglement cost, and relative entropy of entanglement. Our proof constructs an explicit LOCC protocol that converts either of two arbitrary bipartite states into the other at an ebit cost that vanishes as their trace distance tends to zero.
\end{abstract}

\section{Introduction}
Two fundamental tasks in quantum information processing are entanglement distillation and dilution. In the first we are given $n$ copies of a bipartite density matrix and the task is to distill as many Bell states (ebits) as possible using local operations and classical communication (LOCC). The second task reverses this conversion: the goal is to prepare $n$ copies of a fixed bipartite density matrix using LOCC and as few ebits as possible. Allowing for vanishing error, the optimal asymptotic rates, measured in ebits per copy, are called \emph{distillable entanglement}~\cite{bennett96,BBPSSW96,bennett_mixed-state_1996} and \emph{entanglement cost}~\cite{HHT01}, respectively.

To make these operational notions mathematically precise, let $\Phi$ denote a Bell state and $\St(\cH)$ be the set of density matrices on $\cH$. 
Given a bipartite density matrix $\rho_{AB} \in \St(\cH_{AB})$, the distillable entanglement is given by
\begin{align} \label{eq_def-entanglement-distillation}
    E_D(\rho)\coloneqq \lim_{\eps \to 0} \lim_{n \to \infty} \sup \Big \{ \frac{m}{n}:  \inf_{\cD_n \in \LOCC} \frac{1}{2} \norm{ \cD_n(\rho^{\otimes n}) - \Phi^{\otimes m} }_1 \leq \eps \Big\} 
\end{align}
and the entanglement cost by
\begin{align}  \label{eq_def-entanglement-cost}
    E_C(\rho)\coloneqq \lim_{\eps \to 0} \lim_{n \to \infty} \inf \Big \{ \frac{m}{n} : \inf_{\cC_n \in \LOCC} \frac{1}{2} \norm{ \cC_n(\Phi^{\otimes m}) - \rho^{\otimes n} }_1 \leq \eps \Big\}  \, ,
\end{align}
where $\norm{X}_1 \coloneqq \tr[|X|]$ is the trace norm with $|X|\coloneqq\sqrt{X^\dagger X}$.
Entanglement cost features a useful formula~\cite{HHT01}
\begin{align} \label{eq_hayden_formula}
    E_C(\rho) 
    = \lim_{n \to \infty} \frac{1}{n} E_F(\rho^{\otimes n}) 
    = \inf_{n \in \N}\frac{1}{n} E_F(\rho^{\otimes n})  \, ,
\end{align}
where $E_F(\rho)\coloneqq\min_{\{p_i \ket{\psi_i}\}}\{ \sum_i p_i H(\tr_{B}[\proj{\psi_i}]): \rho_{AB}=\sum_i p_i \proj{\psi_i} \}$ denotes the \emph{entanglement of formation}~\cite{bennett_mixed-state_1996} and $H(\sigma)\coloneqq-\tr[\sigma \log \sigma]$ is the von Neumann entropy. 
Note that $E_C$ is different from $E_F$~\cite{hastings09,shor_additivity_04}.
Distillable entanglement has a similar formula~\cite[Theorem~15]{DW05}
\begin{align}\label{eq_winter_formula}
    E_D(\rho) 
    = \lim_{n \to \infty} \frac{1}{n} \sup_{\cP_n \in \LOCC} - H(A^n |B^n)_{\cP_n(\rho^{\otimes n})} 
    =\sup_{n \in \N}  \sup_{\cP_n \in \LOCC} - \frac{1}{n} H(A^n |B^n)_{\cP_n(\rho^{\otimes n})} \, ,
\end{align}
where the second step above follows by Fekete’s lemma~\cite{fekete23}. Due to the complicated structure of the LOCC set~\cite{LOCC14} the formula given in~\cref{eq_winter_formula} is arguably less useful than~\cref{eq_hayden_formula}.

Using~\cref{eq_hayden_formula} it has been shown that the entanglement cost is continuous. More precisely, for two density matrices $\rho$ and $\sigma$ such that $\frac{1}{2}\norm{\rho -\sigma}_1 \leq \eps$ and $\xi=\sqrt{\eps(2-\eps)}$ we have~\cite[Corollary~4]{win16}
\begin{align} \label{eq_cont_EC}
    |E_C(\rho) - E_C(\sigma) | \leq 2\xi \log d + (1+\xi) h\Big(\frac{\xi}{1+\xi} \Big) \, ,
\end{align}
where $h(x)\coloneqq-x\log(x)-(1-x) \log(1-x)$ denotes the binary entropy and $d=\min\{\dim A, \dim B\}$. All logarithms are with base $2$.
\Cref{eq_cont_EC} is particularly useful as it only depends logarithmically on the dimension.

Before this work, distillable entanglement was not known to be in general continuous, let alone to satisfy an explicit continuity bound of the form~\cref{eq_cont_EC}. Over the years, several partial results towards a continuity proof had
been obtained.
From~\cref{eq_winter_formula} it follows that $\rho \mapsto E_D(\rho)$ is lower semicontinuous, since the supremum of continuous functions is lower semicontinuous~\cite[Lemma 2.41]{hitch_book}.
It was shown in~\cite{Vidal02} that the distillable entanglement is continuous on the interior of the set of distillable states.\footnote{This result does not rule out discontinuities of $\rho \mapsto E_D(\rho)$ either at the boundary of the state space, in particular when approaching pure states through mixed states, or at the boundary between distillable and nondistillable states.}

\begin{remark}[Intuitive argument for continuity]
Let $\rho$ and $\sigma$ be two density matrices with trace distance $\delta=\frac{1}{2}\norm{\rho - \sigma}_1$ and $\lambda > \delta$, where $\lambda \coloneqq \lambda^+_{\min}(\rho)$ denotes the smallest non-zero eigenvalue of $\rho$. Then there exists a unitary $U$ such that 
\begin{align}
    \norm{U - \id}_{\infty}\leq\frac{\delta}{\lambda-\delta} 
    \qquad \textnormal{and} \qquad
    \sigma=U\Big( \big(1-\frac{\delta}{\lambda}\big)\rho+\frac{\delta}{\lambda} \tau \Big) U^\dagger \, ,
\end{align}
where $\tau$ is an auxiliary density matrix. This is rigorously justified in~\cref{rmk_existence_unitary}.
Since $U$ is close to the identity, up to local unitaries, it can  be decomposed into a fixed number of two-qubit rotations whose angles tend to zero as $\norm{U-\id}_{\infty}\to0$.\footnote{The inverse function theorem allows to decompose $U$ sufficiently close to the identity into a fixed number of Pauli-string rotations with vanishing angles. Furthermore, each nonlocal Pauli-string rotation is locally
equivalent to a two-qubit rotation with the same angle.}
The protocol in~\cite{SS26Tele} shows that such gates can be implemented with vanishing entanglement cost. This approach proves continuity of $\rho \mapsto E_D(\rho)$, however, the explicit continuity bound depends on $\lambda^+_{\min}(\rho)$. 
To prove the stronger asymptotic continuity with logarithmic dimension dependence (given in~\cref{thm_continuity_ED}), we follow a more tailored proof strategy.
\end{remark}

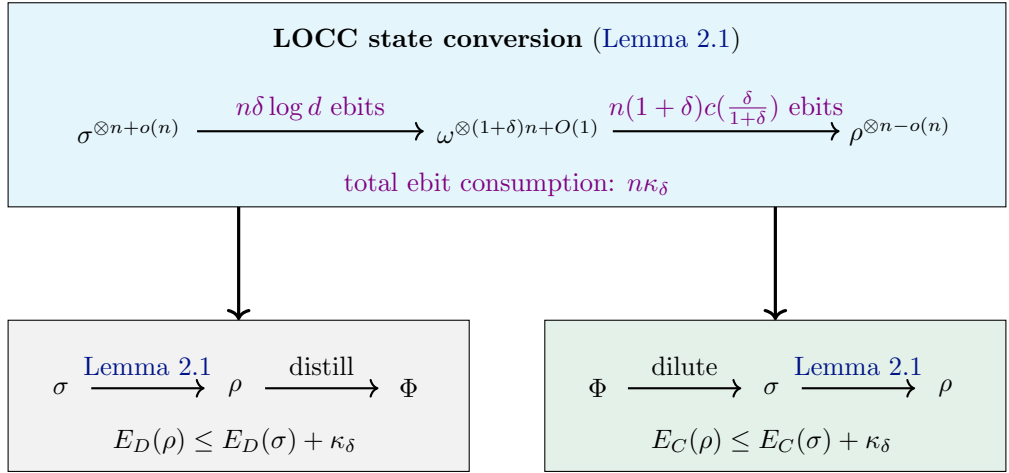
\begin{figure}[!htb]
\centering
\begin{tikzpicture}
\def \x{3};
\def \xx{2.3};

\draw[fill=ProcessBlue!10] (-2.5,-1) rectangle (2*\x+4.2+0.5,1.7);
\node at (\x+3.1-2,1.2) {\textbf{LOCC state conversion} (\cref{lem:state-conversion})};
\node at (\x+3.1-2,-0.7) {{\color{violet}total ebit consumption: $n \kappa_{\delta}$}};

\draw[thick,->] (0,0)--(\x,0);
\node at (-0.9,0) {$\sigma^{\otimes n + o(n)}$};

\node at (\x+1.25,0) {$\omega^{\otimes (1+\delta)n + O(1)}$};
\draw[thick,->] (\x+2.5,0)--(\x+2.5+\x,0);
\node at (2*\x+3.3,0) {$\rho^{\otimes n - o(n)}$};

\node at (0.5*\x,0.3) {{\color{violet}$n \delta \log d$ ebits}};
\node at (1.5*\x+2.5,0.3) {{\color{violet}$n (1+\delta) c(\frac{\delta}{1+\delta}) $ ebits}};


\draw[fill=gray!10] (-2.5,-2.5) rectangle (\x+3.1-2.5,-4.5);
\draw[fill=ForestGreen!10] (2*\x+4.2+0.5,-2.5) rectangle (\x+3.1-1.5,-4.5);
\draw[->,very thick] (-2.5+0.5*\x+1.55,-1) -- (-2.5+0.5*\x+1.55,-2.5);
\draw[->,very thick] (2.35+1.5*\x+0.8,-1) -- (2.35+1.5*\x+0.8,-2.5);

\node at (-2.5+0.7,-2.5-0.9) {$\sigma$};
\node at (-2.5+0.7+\xx,-2.5-0.9) {$\rho$};
\node at (-2.5+0.7+2*\xx,-2.5-0.9) {$\Phi$};
\draw[thick,->] (-2.5+0.7+0.4,-2.5-0.9) -- (-2.5+0.7+\xx-0.4,-2.5-0.9);
\node at (-2.5+0.7+0.5*\xx,-2.5-0.9+0.3) {\cref{lem:state-conversion}};
\draw[thick,->] (-2.5+0.7+\xx+0.4,-2.5-0.9) -- (-2.5+0.7+2*\xx-0.4,-2.5-0.9);
\node at (-2.5+0.7+1.5*\xx,-2.5-0.9+0.3) {distill};
\node at (-2.5+0.7+\xx,-2.5-0.9-0.7) {$E_D(\rho) \leq E_D(\sigma) + \kappa_{\delta} $};

\node at (\x+3.1-1.5+0.7,-2.5-0.9) {$\Phi$};
\node at (\x+3.1-1.5+0.7+\xx,-2.5-0.9) {$\sigma$};
\node at (\x+3.1-1.5+0.7+2*\xx,-2.5-0.9) {$\rho$};
\draw[thick,->] (\x+3.1-1.5+0.7+0.4,-2.5-0.9) -- (\x+3.1-1.5+0.7+\xx-0.4,-2.5-0.9);
\node at (\x+3.1-1.5+0.7+0.5*\xx,-2.5-0.9+0.3) {dilute};
\draw[thick,->] (\x+3.1-1.5+0.7+\xx+0.4,-2.5-0.9) -- (\x+3.1-1.5+0.7+2*\xx-0.4,-2.5-0.9);
\node at (\x+3.1-1.5+0.7+1.5*\xx,-2.5-0.9+0.3) {\cref{lem:state-conversion}};
\node at (\x+3.1-1.5+0.7+\xx,-2.5-0.9-0.7) {$E_C(\rho) \leq E_C(\sigma) + \kappa_{\delta} $};
\end{tikzpicture}
\vspace{-7mm}
\caption{The key element of the proof is the ``LOCC state conversion lemma,'' which converts $n+o(n)$ copies of $\sigma$, using $n\kappa_{\delta}+o(n)$ ebits, into an approximation of $n-o(n)$ copies of $\rho$ with vanishing trace-distance error, where $\kappa_{\delta}$ is given in~\cref{eq_kappa}. Concatenating this conversion with an entanglement distillation protocol yields $nE_D(\rho)-o(n)$ ebits with vanishing error. \Cref{lem_basic_but_powerful} then implies $E_D(\rho)\leq E_D(\sigma)+\kappa_{\delta}$. Since $\kappa_{\delta}$ is unchanged when $\rho$ and $\sigma$ are interchanged, we obtain $|E_D(\rho)-E_D(\sigma)|\leq\kappa_{\delta}$, establishing continuity.
A similar argument proves continuity of $E_C$, as explained in~\cref{rmk_EC_continuity}.}
\label{fig_overview_proof}
\end{figure}

\begin{theorem} \label{thm_continuity_ED}
Let $\eps \in [0,\frac{1}{16})$, $\rho,\sigma \in \St(\cH_{AB})$ such that $\frac{1}{2} \norm{\rho-\sigma}_1 \leq \eps$, and $d = \min\{\dim A, \dim B\}$. Then
    \begin{align} \label{eq_cont_main}
        |E_D(\rho) - E_D(\sigma)| \leq 4 \sqrt{\eps} \log d   + 4 h(\sqrt{\eps})\, .
    \end{align}   
\end{theorem}
The proof is given in~\cref{sec_proof} and its high level idea is summarized in~\cref{fig_overview_proof}. The proof uses the operational definition of distillable entanglement and is therefore conceptually different from the continuity proof for entanglement cost which employs the variational formula~\cref{eq_hayden_formula}.
For a Hermitian matrix $H$ let \smash{$H_{+}\coloneqq\frac{|H|+H}{2}$} and \smash{$H_{-}\coloneqq\frac{|H|-H}{2}$} be its positive and negative parts, respectively, and let $\delta\coloneqq\delta(\rho,\sigma)\coloneqq\frac{1}{2} \norm{\rho - \sigma }_1$ be the trace distance. 
To understand the proof of~\cref{thm_continuity_ED}, let \smash{$\gamma_{+}\coloneqq \frac{(\rho- \sigma)_{+}}{\delta}$} and \smash{$\gamma_{-}\coloneqq\frac{(\rho-\sigma)_{-}}{\delta}$}, which are both density matrices as they are by definition clearly positive semidefinite and $\tr[\gamma_{+}]=\tr[\gamma_{-}]=1$, because $(\rho- \sigma)_{+}$ and $(\rho- \sigma)_{-}$ have orthogonal support and therefore $\tr[(\rho- \sigma)_{+}]=\tr[(\rho- \sigma)_{-}]=\delta$.
Let 
\begin{align}
\omega \coloneqq \Big( 1 - \frac{\delta}{1+\delta} \Big)\rho + \frac{\delta}{1+\delta} \gamma_{-}
= \frac{\rho + \delta \gamma_-}{1+\delta}
= \frac{\sigma + \delta \gamma_+}{1+\delta}
=\Big( 1 - \frac{\delta}{1+\delta} \Big)\sigma + \frac{\delta}{1+\delta} \gamma_{+} \, . \label{eq_state_omega}
\end{align}
As visualized in~\cref{fig_overview_proof}, the density matrix $\omega$ will play a crucial role.\footnote{The proofs in~\cite{winter05} use a similar decomposition as~\cref{eq_state_omega}.}
Finally, let 
\begin{align} \label{eq_kappa}
 \kappa_{\delta}\coloneqq\delta\log d+(1+\delta)c\!\left(\frac{\delta}{1+\delta}\right) \, ,
\end{align}
where $c:[0,1] \to \R_+$ is given by
\begin{equation} \label{eq_c_function}
        c(x) \coloneqq \pi\sqrt{x}\log\frac{4\ee d}{\sqrt x} + \frac{3}{2}x\log d - \frac{3}{4}x\log\frac{16\ee}{x}  \, .
\end{equation}
Note that $c(0)=0$ and $\lim_{\delta \to 0} \kappa_{\delta} = 0$. 

\begin{remark} \label{rmk_EC_continuity}
\Cref{eq_cont_main} in~\cref{thm_continuity_ED} also holds with $E_D$ replaced by $E_C$.
The idea is visualized in~\cref{fig_overview_proof} and proved in~\cref{cor:cost}.
The continuity bound given in~\cref{eq_cont_EC} has a similar scaling, however, is slightly tighter.
\end{remark}

\section{Proof of~\cref{thm_continuity_ED}} \label{sec_proof}
\begin{figure}[!htb]
\centering
\begin{tikzpicture}
\def \xb{3.6};
\def \yb{1.6};

\def \x{0.3};
\def \y{0.4};

\def \xx{3.8};

\draw [] (0,0) rectangle (\xb,\yb);
\node at (0.5*\xb,0.8*\yb) {\Cref{thm_continuity_ED}};
\node at (0.5*\xb,0.45*\yb) {Asymptotic continuity};
\node at (0.5*\xb,0.2*\yb) {of $\rho \mapsto E_D(\rho)$};

\draw [] (-\x,-\y) rectangle (-\x-\xb,-\y-\yb);
\node at (-\x-0.5*\xb,-\y-0.2*\yb) {\Cref{lem:state-conversion}};
\node at (-\x-0.5*\xb,-\y-0.57*\yb) {LOCC state};
\node at (-\x-0.5*\xb,-\y-0.8*\yb) {conversion};
\draw[->,thick] (-\x-0.5*\xb,-\y) -- (0,0.5*\yb);

\draw [] (\xb+\x,-\y) rectangle (\x+2*\xb,-\y-\yb);
\node at (\x+1.5*\xb,-\y-0.2*\yb) {\Cref{lem_basic_but_powerful}};
\node at (\x+1.5*\xb,-\y-0.57*\yb) {Ebit-assisted};
\node at (\x+1.5*\xb,-\y-0.8*\yb) {distillation bound};
\draw[->,thick] (\x+1.5*\xb,-\y) -- (\xb,0.5*\yb);

\draw [] (-\x,-2*\y-\yb) rectangle (-\x-\xb,-2*\y-2*\yb);
\node at (-\x-0.5*\xb,-2*\y-1.2*\yb) {\Cref{lem_main}};
\node at (-\x-0.5*\xb,-2*\y-1.57*\yb) {Measurement};
\node at (-\x-0.5*\xb,-2*\y-1.8*\yb) {(LOCC)};

\draw[->,thick] (-\x-0.5*\xb,-2*\y-\yb) -- (-\x-0.5*\xb,-\y-\yb);

\draw [] (-\x,-3*\y-2*\yb) rectangle (-\x-\xb,-3*\y-3*\yb);
\node at (-\x-0.5*\xb,-3*\y-2.2*\yb) {\Cref{lem:algo}};
\node at (-\x-0.5*\xb,-3*\y-2.57*\yb) {Identification of};
\node at (-\x-0.5*\xb,-3*\y-2.8*\yb) {outcomes (LOCC)};

\draw[->,thick] (-\x-0.5*\xb,-3*\y-2*\yb) -- (-\x-0.5*\xb,-2*\y-2*\yb);

\draw [] (-\x-\xx,-3*\y-2*\yb) rectangle (-\x-\xb-\xx,-3*\y-3*\yb);
\node at (-\x-0.5*\xb-\xx,-3*\y-2.2*\yb) {\Cref{lem:oracle}};
\node at (-\x-0.5*\xb-\xx,-3*\y-2.57*\yb) {Controlled phase};
\node at (-\x-0.5*\xb-\xx,-3*\y-2.8*\yb) {oracle (LOCC)};

\draw[->,thick] (-\x-0.5*\xb-\xx,-3*\y-2*\yb) -- (-\x-\xb,-2*\y-1.5*\yb);

\draw [] (-\x+\xx,-3*\y-2*\yb) rectangle (-\x-\xb+\xx,-3*\y-3*\yb);
\node at (-\x-0.5*\xb+\xx,-3*\y-2.2*\yb) {\Cref{lem:approx}};
\node at (-\x-0.5*\xb+\xx,-3*\y-2.57*\yb) {Error};
\node at (-\x-0.5*\xb+\xx,-3*\y-2.8*\yb) {propagation};

\draw[->,thick] (-\x-0.5*\xb+\xx,-3*\y-2*\yb) -- (-\x,-2*\y-1.5*\yb);
\end{tikzpicture}
\vspace{-7mm}
\caption{Relations of all lemmas used to prove the assertion of~\cref{thm_continuity_ED}.}
\label{fig_structure}
\end{figure}
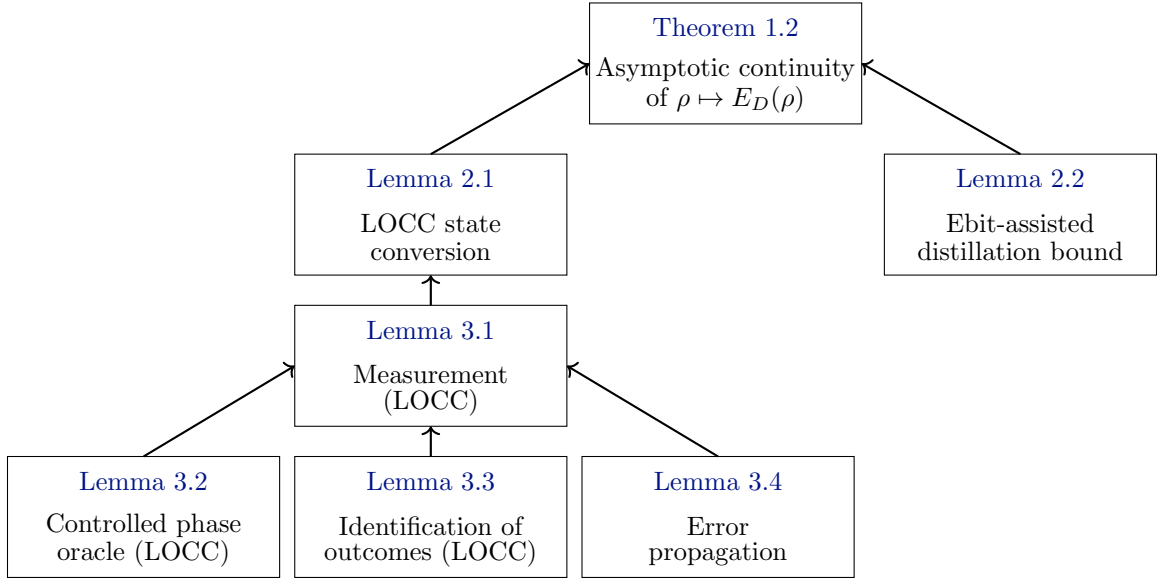

The high-level idea of the proof is summarized in~\cref{fig_overview_proof}. The main step is the ``LOCC state conversion lemma" that transforms $\sigma$ asymptotically to $\rho$ with an ebit cost controlled by $\delta$.
The logical relations of all the lemmas used are visualized in~\cref{fig_structure}.

\begin{lemma}[LOCC state conversion]\label{lem:state-conversion}
    Let $\rho,\sigma\in \St(\cH_{AB})$ such that $\delta\coloneqq\frac{1}{2}\|\rho-\sigma\|_1<\frac1{16}$, $d\coloneqq\min\{\dim A,\dim B\}$ and $\kappa_{\delta}$ as defined in~\cref{eq_kappa}. There exist integers $N_n=n+\lceil n^{2/3}\rceil+1$, $M_n=n- \lceil n^{2/3}\rceil$ and LOCC channels $(\mathcal P_n)_{n \in \N}$ such that
    \begin{equation}
        \norm{\cP_n\!\left(\sigma^{\otimes N_n}\otimes\Phi^{\otimes k_n}\right)-\rho^{\otimes M_n}}_1 =o(1) \, ,
    \end{equation}
    where $k_n = n \kappa_{\delta} + o(n)$.
\end{lemma}
\begin{proof}
The proof consists of two steps as visualized in~\cref{fig_overview_proof}.

\textbf{Step 1: Prepare the intermediate state $\omega$.}
Recalling the decomposition given in~\cref{eq_state_omega}, by the asymptotic mixing lemma~\cite[Eq.~(21)]{Vidal02}, there exists an LOCC protocol that uses $N_n$ copies of $\sigma$ and $\delta n+o(n)$ copies of $\gamma_+$ to produce a state $\tilde\omega_n$ satisfying
    \begin{equation}
        \delta(\tilde\omega_n,\omega^{\otimes \lceil(1+\delta )n\rceil})=o(1) \, .
    \end{equation}
    Preparing the auxiliary copies of $\gamma_+$ by teleporting the smaller subsystem requires $\delta n\log d+o(n)$ ebits.  We recall the argument here for completeness.
    Using~\cref{eq_state_omega} sample iid random variables $(Z_i)_{i \in [\lceil(1+\delta)n\rceil]} \sim \textnormal{Bernoulli}(\frac{\delta}{1+\delta})$ such that
    \begin{align}
        \omega^{\otimes\lceil(1+\delta)n\rceil} = \E\Big[\bigotimes_{i=1}^{\lceil(1+\delta)n\rceil} \begin{cases} \sigma , & Z_i =0 \\
        \gamma_+, & Z_i = 1 \end{cases}\Big] \, .
    \end{align}
    We prepare $\omega^{\otimes\lceil(1+\delta)n\rceil}$ by using copies of $\sigma$ and ebits (for preparing $\gamma_+$). In the following we quantify how many copies of $\sigma$ and  how many ebits we require.
    Let $Z\coloneqq\sum_{i=1}^{\lceil(1+\delta)n\rceil} Z_i$ and note that $Z\sim\operatorname{Bin}(\lceil(1+\delta)n\rceil,\frac{\delta}{1+\delta})$ and $\E[Z]=\frac{\delta}{1+\delta}\lceil(1+\delta)n\rceil$.
    Hoeffding's inequality gives
    \begin{align}\label{eq_hoeffding}
        \PP\Big(\big|\,Z-\E[Z]\,\big|>n^{2/3}\Big) \leq 2\exp\!\left(-\frac{2n^{4/3}}{\lceil(1+\delta)n\rceil}\right)=o(1)\, .
    \end{align}
    If $\big|\,Z-\E[Z]\,\big|\leq n^{2/3}$, the number of $\sigma$ positions is at most
    \begin{align}
        \lceil(1+\delta)n\rceil-Z
        \leq\frac{\lceil(1+\delta)n\rceil}{1+\delta}+n^{2/3}
        \leq n+n^{2/3}+1\, .
    \end{align}
    Hence,
    \begin{align}
        N_n\coloneqq n+\lceil n^{2/3}\rceil+1=n+O(n^{2/3})
    \end{align}
    copies of $\sigma$ suffice.
    Under the same condition, $Z\leq n\delta+n^{2/3}+1$, and therefore we need
    \begin{align}
        e_n \coloneqq \left\lceil\log d\big\lceil n\delta+n^{2/3}+1\big\rceil\right\rceil
        \leq n\delta\log d+o(n)
    \end{align}
    ebits to prepare the auxiliary copies of $\gamma_+$.
    If $\big|\,Z-\E[Z]\,\big|>n^{2/3}$, the protocol instead outputs a fixed product state.

    Let $\tilde\omega_n$ denote the state produced using $N_n$ copies of $\sigma$ and $e_n$ ebits. The preparation error satisfies
    \begin{align}
        \alpha_n \coloneqq \delta \big( \omega^{\otimes\lceil(1+\delta)n\rceil} , \tilde\omega_n\big)
        \leq\PP\Big(\big|\,Z-\E[Z]\,\big|>n^{2/3}\Big) \overset{\textnormal{\Cshref{eq_hoeffding}}}{=}o(1)\, .
    \end{align}
    We next extend the ideal and the actual output state by a qubit register $C$ on the party with the larger local dimension, i.e. $\omega^{\mathrm{anc}}\coloneqq\omega\otimes \proj{0}_C$ and $\tilde \omega_n^{\mathrm{anc}} \coloneqq \tilde\omega_n \otimes (\proj{0})^{\otimes\lceil(1+\delta)n \rceil}$. 
    We note that
    \begin{align} \label{eq_alpha_n}
    \delta\big((\omega^{\mathrm{anc}})^{\otimes\lceil(1+\delta)n\rceil}, \tilde \omega_n^{\mathrm{anc}} \big) = \alpha_n \, .
    \end{align}
\textbf{Step 2: Extract the target state $\rho$.}
We now use the decomposition of $\omega$ (see~\cref{eq_state_omega})
to extract $\rho$ from $\omega$. 
\Cref{lem:bhatia}, applied to $(1-\frac{\delta}{1+\delta})\rho\leq\omega$, gives
a projector $P$ on $\cH_{ABC}$ such that
\begin{align}\label{eq_mat_from_lemma}
    (\id_{AB}\otimes\langle0|_C)P\omega^{\mathrm{anc}}P (\id_{AB}\otimes|0\rangle_C) = \Big(1-\frac{\delta}{1+\delta}\Big)\rho 
    \qquad \textnormal{and} \qquad
    \tr[(\id-P)\omega^{\mathrm{anc}}]\leq \frac{\delta}{1+\delta} \, .
\end{align}
Let
\begin{align}
    \cM(X)\coloneqq P X P \otimes \proj{0} + (\id-P) X (\id - P)\otimes \proj{1} 
\end{align}
and $\cM_n=\cM^{\otimes\lceil(1+\delta)n\rceil}$.
\Cref{lem_main} gives an LOCC channel $\tilde \cM_n$ consuming $n(1+\delta)c(\frac{\delta}{1+\delta})+o(n)$ ebits, whose error on the ideal input satisfies
\begin{align} \label{eq_beta_n}
 \beta_n \coloneqq \delta\left(\tilde \cM_n\big((\omega^{\mathrm{anc}})^{\otimes\lceil(1+\delta)n\rceil}\big), \cM_n\big((\omega^{\mathrm{anc}})^{\otimes\lceil(1+\delta)n\rceil}\big) \right) = o(1) \, .
\end{align}
Using the triangle inequality~\cite[Lemma~9.1.8]{wilde_book} and contractivity~\cite[Exercise~9.1.8]{wilde_book} of the trace distance we find
\begin{align}
 \delta\left( \tilde \cM_n(\tilde \omega_n^{\mathrm{anc}}), \cM_n\big((\omega^{\mathrm{anc}})^{\otimes\lceil(1+\delta)n\rceil}\big)\right) \overset{\textnormal{\Cshref{eq_alpha_n,eq_beta_n}}}{\leq}\alpha_n+\beta_n \, . \label{eq_alpha_and_beta}
\end{align}

Choose an arbitrary separable state $\vartheta \in \St(\cH_{AB})$ (for instance the maximally mixed state) and define the LOCC channel $\mathcal R$ as follows: Read the classical bit $b$ given by $\cM$, and measure $C$ in the computational basis to obtain
$z\in\{0,1\}$. If $b=z=0$, retain the postmeasurement $AB$ state and set $Y=1$; otherwise replace $AB$ by $\vartheta$ and set $Y=0$.
Share $Y$ between the parties and discard $C$ and the old outcome records.
\Cref{eq_mat_from_lemma} gives
\begin{align}
 (\cR\circ\cM)(\omega^{\mathrm{anc}}) = \Big(1-\frac{\delta}{1+\delta}\Big)\rho\otimes\proj{1}_Y + \frac{\delta}{1+\delta}\vartheta\otimes\proj{0}_Y\, . \label{eq:filter-flagged-output}
\end{align}
Let
\begin{align}
    M_n\coloneqq n-\lceil n^{2/3}\rceil=n-O(n^{2/3})\, .
\end{align}
For $y\in\{0,1\}^{\lceil(1+\delta)n\rceil}$ with $|y|\geq M_n$, let $\cJ_{M_n}(y)$ be the set of its first $M_n$ positions equal to one. Define the LOCC channel 
\begin{align}
 \mathcal S_n\bigl(X\otimes|y\rangle\langle y|\bigr)
 =\begin{cases}
   \tr_{\cJ_{M_n}(y)^c}X,& |y|\geq M_n,\\
   \tr[X]\,\vartheta^{\otimes M_n},& |y|<M_n,
  \end{cases}
\end{align}
where $\cJ_{M_n}(y)^c$ is the complement of $\cJ_{M_n}(y)$.
Hoeffding's inequality gives 
\begin{align} \label{eq_hoeffding_binomial}
    \gamma_n \coloneqq \PP\Big(\operatorname{Bin}\Big(\lceil(1+\delta)n\rceil,1-\frac{\delta}{1+\delta}\Big)<M_n\Big) \leq\exp\!\left(-\frac{2n^{4/3}}{\lceil(1+\delta)n\rceil}\right)=o(1)\, .
\end{align}
\Cref{eq:filter-flagged-output} gives
\begin{align}
    (\cS_n\circ\cR^{\otimes\lceil(1+\delta)n\rceil}\circ\cM_n) \big((\omega^{\mathrm{anc}})^{\otimes\lceil(1+\delta)n\rceil}\big) = (1-\gamma_n)\rho^{\otimes M_n}+\gamma_n\vartheta^{\otimes M_n} \, . \label{eq_almost_done_DS}
\end{align}
Observe that 
\begin{align}
    \delta\big( \cS_n \circ \cR^{\otimes\lceil(1+\delta)n\rceil} \circ \tilde \cM_n(\tilde \omega_n^{\mathrm{anc}}) , \, & \cS_n \circ \cR^{\otimes\lceil(1+\delta)n\rceil} \circ \cM_n((\omega^{\mathrm{anc}})^{\otimes\lceil(1+\delta)n\rceil}) \big) \\
    \overset{\textnormal{contractivity}}&{\leq} \delta\big( \tilde \cM_n(\tilde \omega_n^{\mathrm{anc}}) ,  \cM_n((\omega^{\mathrm{anc}})^{\otimes\lceil(1+\delta)n\rceil}) \big) \\
    \overset{\textnormal{\Cshref{eq_alpha_and_beta}}}&{\leq} \alpha_n + \beta_n \label{eq_step_1_done} 
\end{align}
and
\begin{align}
    \delta\big( \cS_n \circ \cR^{\otimes\lceil(1+\delta)n\rceil} \circ \cM_n((\omega^{\mathrm{anc}})^{\otimes\lceil(1+\delta)n\rceil}) , \rho^{\otimes M_n} \big)
    \overset{\textnormal{\Cshref{eq_almost_done_DS}}}&{\leq} \delta\big( (1-\gamma_n) \rho^{\otimes M_n} + \gamma_n \, \vartheta^{\otimes M_n}, \rho^{\otimes M_n} \big) \\
    \overset{\textnormal{convexity}}&{\leq} \gamma_n \delta( \vartheta^{\otimes M_n}, \rho^{\otimes M_n}) \\
    &\leq \gamma_n \, , \label{eq_step_2_done} 
\end{align}
where the convexity of the trace distance is explained in~\cite[Exercise~9.1.9]{wilde_book}.
Putting things together yields 
\begin{align}
    &\hspace{-20mm}\delta\Big(\cS_n\circ\cR^{\otimes\lceil(1+\delta)n\rceil} \circ\tilde \cM_n (\tilde\omega_n^{\mathrm{anc}}), \rho^{\otimes M_n}\Big) \nonumber \\
    \overset{\textnormal{triangle}}&{\leq} \delta\big( \cS_n \circ \cR^{\otimes\lceil(1+\delta)n\rceil} \circ \tilde \cM_n(\tilde \omega_n^{\mathrm{anc}}) , \cS_n \circ \cR^{\otimes\lceil(1+\delta)n\rceil} \circ \cM_n((\omega^{\mathrm{anc}})^{\otimes\lceil(1+\delta)n\rceil}) \big) \nonumber \\
    &\hspace{40mm}+ \delta\big( \cS_n \circ \cR^{\otimes\lceil(1+\delta)n\rceil} \circ \cM_n((\omega^{\mathrm{anc}})^{\otimes\lceil(1+\delta)n\rceil}) , \rho^{\otimes M_n} \big) \\
    \overset{\textnormal{\Cshref{eq_step_1_done,eq_step_2_done}}}&{\leq} \alpha_n + \beta_n + \gamma_n   \\
    &= o(1) \, . \label{eq_essss_done}
\end{align}


Combining steps 1 and 2 mentioned above, we have constructed an LOCC protocol $\cP_n$ that converts $N_n=n+O(n^{2/3})$ copies of $\sigma$ into $M_n=n-O(n^{2/3})$ copies of $\rho$, with vanishing trace distance error $\alpha_n+\beta_n+\gamma_n$, using
\begin{align}
 n\delta\log d+n(1+\delta)c\Big(\frac{\delta}{1+\delta}\Big)+o(n) \overset{\textnormal{\Cshref{eq_kappa}}}{=} n\kappa_{\delta}+o(n) =: k_n
\end{align} 
ebits as visualized in~\cref{fig_overview_proof}. This proves the lemma.
\end{proof}

\begin{lemma} \label{lem_basic_but_powerful}
    Any LOCC protocol that consumes $n$ copies of a density matrix $\rho \in \St(\cH_{AB})$ and $n c+o(n)$ ebits, and that produces $n d- o(n)$ ebits with vanishing trace-distance error has to fulfill
\begin{equation} \label{eq_basic}
    d\leq E_D(\rho)+c \, ,
\end{equation}
where $d$ and $c$ are constants independent of $n$.
\end{lemma}
\begin{proof}
Let $k_n=nc+o(n)$ be the actual integer number of ebits
required by the assisted protocol, and fix $0<r<E_D(\rho)$.
Set
\begin{equation}
    s_n\coloneqq\left\lceil\frac{k_n+\sqrt n}{r}\right\rceil
        =\frac{nc}{r}+o(n)\, .
\end{equation}
Since $s_n\to\infty$ and $r<E_D(\rho)$, we can distill $\lfloor rs_n\rfloor$ ebits from $s_n$ copies of $\rho$ with vanishing trace-distance error. Moreover, $\lfloor rs_n\rfloor\geq k_n$, so discarding any surplus gives the $k_n$ ebits required by the assisted protocol.
Concatenating this preparation with the assisted protocol gives an unassisted protocol consuming $n+s_n$ copies of $\rho$ and producing $nd-o(n)$ ebits. Its error vanishes by the triangle inequality and contractivity of the trace distance.
The resulting asymptotic rate is
\begin{equation}
    \lim_{n\to\infty}\frac{nd-o(n)}{n+s_n}
    =\frac{d}{1+c/r} \, ,
\end{equation}
which implies
\begin{equation}
    E_D(\rho)\geq\frac{d}{1+c/r} \, .
\end{equation}
Sending $r\uparrow E_D(\rho)$ gives~\cref{eq_basic}.

    In the case $E_D(\rho) = 0$ we need to show $d \leq c$. Assume, for contradiction, that an LOCC protocol $\Lambda$ achieves $d>c$. This protocol maps $\rho^{\otimes n} \otimes \Phi^{\otimes n(c+o(1))}$ to a state $\sigma_n$ which is close to $\Phi^{\otimes n(d-o(1))}$. This error tends to zero, so $\tr[\Phi^{\otimes n(d-o(1))}\sigma_n] \to 1$. Since $d>c$, for sufficiently large $n$ we have
    \begin{equation}
        \tr[\Phi^{\otimes n(d-o(1))}\sigma_n] \, 2^{n(d-o(1))} > 2^{n(c+o(1))} \, .
    \end{equation}
    Now consider the separable state
    \begin{equation}
        \eta = 2^{-n(c+o(1))} \sum_i \ket{ii}\bra{ii}_{AB} \, .
    \end{equation}
    For this state we have the inequality
    \begin{equation}
        \eta \geq 2^{-n(c+o(1))} \, \Phi^{\otimes n(c+o(1))} \, .
    \end{equation}
    If we apply the same LOCC channel to this input, we get $\sigma_n' = \Lambda(\rho^{\otimes n}\otimes \eta)$ and using positivity and linearity, this state satisfies $\sigma'_n \geq  2^{-n(c+o(1))} \,\sigma_n$. Therefore 
    \begin{equation}
        \tr[\sigma'_n \Phi^{\otimes n(d-o(1))}] \geq 2^{-n(c+o(1))}\tr[\Phi^{\otimes n(d-o(1))}\sigma_n] > 2^{-n(d-o(1))} \, .
    \end{equation}
    This however contradicts $E_D(\rho)=0$, since a state whose overlap with a maximally entangled state of Schmidt rank $D$ exceeds $1/D$ is distillable~\cite{Horodecki-Horodecki}.
\end{proof}

\begin{proof}[Proof of~\cref{thm_continuity_ED}]
We conclude the proof by applying entanglement distillation to the converted state (see~\cref{fig_overview_proof}). Fix $0<r<E_D(\rho)$  and let $N_n$, $M_n$, and $k_n$ be as before. Starting with $N_n$ copies of $\sigma$ and $k_n$ Bell states, the protocol from~\cref{lem:state-conversion} produces an approximation to $\rho^{\otimes M_n}$ with vanishing trace-distance error. Concatenating this protocol with an entanglement distillation protocol $\cD_n$ for $\rho$ at rate $r$ produces an approximation to $\lfloor M_n r \rfloor$ Bell states.
\begin{align}
    \delta\big(\cD_n \!\circ\! \cP_n(\sigma^{\otimes N_n} \otimes \Phi^{\otimes k_n}), \Phi^{\otimes M_n r} \big)
    \!\overset{\textnormal{triangle}}&{\leq}\! \delta\big(\cD_n \!\circ\! \cP_n(\sigma^{\otimes N_n} \otimes \Phi^{\otimes k_n}), \cD_n(\rho^{\otimes M_n}) \big) \!+\! \delta\big( \cD_n(\rho^{\otimes M_n}),\Phi^{\otimes M_n r}\big) \\
    \overset{\textnormal{contraction}}&{\leq}\delta(\cP_n(\sigma^{\otimes N_n} \otimes \Phi^{\otimes k_n}), \rho^{\otimes M_n}) + o(1) \\
    \overset{\textnormal{\Cshref{lem:state-conversion}}}&{=} o(1) \, .
\end{align}
Therefore using~\cref{lem_basic_but_powerful} gives $r \leq E_D(\sigma) + \kappa_{\delta}$ and letting $r \uparrow E_D(\rho)$ together with~\cref{eq_kappa} results in
\begin{align} \label{eq_one_sided}
    E_D(\rho) - E_D(\sigma) \leq \delta \log d + (1+\delta) c\Big(\frac{\delta}{1+\delta}\Big) \, .
\end{align}
Interchanging $\rho$ and $\sigma$ and using $0 \leq E_D \leq \log d$ leads to
\begin{align} 
    |E_D(\rho) - E_D(\sigma)| 
    \leq \min\Big\{\log d, \delta \log d + (1+\delta) c\Big(\frac{\delta}{1+\delta}\Big)\Big \} 
    \overset{\textnormal{\Cshref{lem_scalar}}}{\leq} 4 \sqrt{\delta} \log d + 4 h(\sqrt{\delta})\, .
\end{align}
\end{proof}

\begin{corollary} \label{cor:cost}
Let $\eps \in [0,\frac{1}{16})$, $\rho,\sigma \in \St(\cH_{AB})$ such that $\frac{1}{2} \norm{\rho-\sigma}_1 \leq \eps$, and $d = \min\{\dim A, \dim B\}$. Then
    \begin{align} 
        |E_C(\rho) - E_C(\sigma)| \leq 4 \sqrt{\eps} \log d   + 4 h(\sqrt{\eps})\, .
    \end{align}   
\end{corollary}
\begin{proof}
    Fix $r>E_C(\sigma)$ and let $N_n$, $M_n$, and $k_n$ be as before. Starting with $rN_n+k_n+o(n)$ ebits, spend $rN_n+o(n)$ ebits to prepare an approximation to $\sigma^{\otimes N_n}$ and reserve $k_n$ ebits for the state conversion. Applying~\cref{lem:state-conversion} gives an approximation to $\rho^{\otimes M_n}$ with vanishing trace-distance error. Since
    \begin{align}
        rN_n+k_n+o(n) \leq n(r+\kappa_\delta)+o(n)
    \end{align}
    and $M_n=n-o(n)$, this is a dilution protocol for $\rho$ with asymptotic cost at most $r+\kappa_\delta$ ebits per output copy. Since $E_C(\rho)$ is the infimum of achievable dilution rates, $E_C(\rho)\leq r+\kappa_{\delta}$.
    Let $r\downarrow E_C(\sigma)$, interchange $\rho$ and $\sigma$, apply $0\leq E_C\leq\log d$, and~\cref{lem_scalar} to obtain the claim. 
\end{proof}

\section{Ingredients for the proof of~\cref{lem:state-conversion}}
In this section, we state and prove the lemmas needed for the proof of~\cref{lem:state-conversion} as visualized in~\cref{fig_structure}.
For a projector $P$ on $\cH_{AB}$, define $P_0\coloneqq P$, $P_1\coloneqq\id-P=:P^\perp$, and
\begin{equation} \label{eq_projector}
    \Pi_z\coloneqq\bigotimes_{i=1}^n P_{z_i} \qquad \textnormal{for} \qquad  z\in\{0,1\}^n \, .
\end{equation}
The corresponding measurement channel is given by
\begin{equation}
 \cM_n(X)=\sum_z\Pi_zX\Pi_z \otimes\ket{z}\bra{z}_Z \, ,
 \label{eq:instrument}
\end{equation}
where the classical register $Z$ stores the measurement outcome. This value can be freely shared between the parties.
Our goal is to implement this channel applied to some state $\omega^{\otimes n}$ using LOCC and as few ebits as possible. 
A direct implementation teleports the smaller subsystem of each copy to the other party, performs the measurement there, and teleports it back, using $O(n)$ ebits. But if one of the outcomes is rare, we can perform a collective strategy that searches for the few positions where the less likely outcomes occur.

\begin{lemma} \label{lem_main}
    Let $p \in [0,\frac{1}{16})$, $\omega\in \St (\cH_{AB})$, and $P$ be a projector satisfying $\tr[(\id - P)\omega] \leq p$. Then using LOCC and
    $nc(p)+o(n)$ ebits, one can implement the channel $\widetilde{\cM}_n$ satisfying
    \begin{equation}
        \norm{\widetilde{\cM}_n(\omega^{\otimes n})- \cM_n(\omega^{\otimes n})}_1 = o(1)\, ,
    \end{equation}
    where the function $c(\cdot)$ is defined in~\cref{eq_c_function} and $\cM_n$ in~\cref{eq:instrument}.
\end{lemma}
\begin{proof}
    The case $p=0$ is straightforward and hence assume in the following $p>0$.
    Consider a purification $\ket{\Omega}_{A^nB^nR}$ of $\omega^{\otimes n}$ and use~\cref{eq_projector} to write it as 
    \begin{equation}
        \ket{\Omega}_{A^nB^nR} = \sum_z \ket{\Omega_z} \qquad \textnormal{for} \qquad \ket{\Omega_z} = (\Pi_z \otimes \id_R) \ket{\Omega}_{A^nB^nR} \, .
    \end{equation}
    If we could do
    \begin{equation}
         \ket{\Omega}_{A^nB^nR} \mapsto \sum_z \ket{\Omega_z} \otimes \ket{z}_Z\ket{\xi_z}_E \, ,
    \end{equation}
    where $ \ket{z}_Z$ encodes the bit string $z$ and $\ket{\xi_z}_E$ a perfect copy in the environment, then tracing out $E$ (using orthogonality of $\ket{\xi_z}_E$) and $R$ yields $\cM_n(\omega^{\otimes n})$.
    Implementing $\cM_n$ can therefore be seen as attempting to learn the different strings $z$ on each subspace.

    On a component $\ket{\Omega_z}$, applying the projector $P^\perp$ on the $i$-th subsystem results in
    \begin{equation} \label{eq_Pi_z}
        P_i^\perp \ket{\Omega_z} = z_i \ket{\Omega_z}
    \end{equation}
    and therefore we have $(\id-2P_i^\perp)$ acting like a phase oracle for the bit $z_i$ on this subspace, i.e.
    \begin{equation}
        (\id-2P_i^\perp)\ket{\Omega_z} 
        \overset{\textnormal{\Cshref{eq_Pi_z}}}{=} (1-2z_i)\ket{\Omega_z} 
        = (-1)^{z_i}\ket{\Omega_z} \, ,
    \end{equation}
    where the final step uses $z_i \in \{0,1\}$.
    The key idea to implement $\cM_n$ is to make use of this phase oracle to implement Grover's algorithm to identify $z$. This is done by searching for all positions for which $z_i=1$. The actual algorithm then takes multiple random short lists and implements the usual Grover algorithm on these. A similar approach has been taken in~\cite[Corollary~6.5]{AaronsonAmbainis2005}.
    
    We give the ebit cost of a single oracle call in~\cref{lem:oracle} and the analysis of the full algorithm in~\cref{lem:algo}. Its application has the form
    \begin{equation} \label{eq_operation}
        \ket{\Omega} \mapsto \sum_z \ket{\Omega_z} \otimes \ket{v_z}_{ZE}
    \end{equation}
    where the register $Z$ stores the final result of the algorithm and the environment $E$ contains its records, including a copy of this result. By \cref{lem:algo}, the protocol uses $nc(p)+o(n)$ ebits and reports the correct string $z$ with probability at least $1-\delta_n$ on every component $\ket{\Omega_z}$ with $|z|\leq r_n$, where $\delta_n\to0$.
    
    It remains to bound the total weight of the other components. Since the input state is $\omega^{\otimes n}$ and $\tr[P^\perp\omega]\leq p$, the number of ones in $z$ is binomially distributed with mean at most $pn$. Hoeffding's inequality therefore gives
    \begin{equation}
        \sum_{|z|>r_n}\tr[\Pi_z\omega^{\otimes n}]
        \leq \ee^{-2n^{1/3}},
    \end{equation}
    where we used $r_n=\lceil pn+n^{2/3}\rceil$.
    The average error over the components $\ket{\Omega_z}$ is thus at most $\delta_n+\ee^{-2n^{1/3}}$.
    Writing $\widetilde{\cM}_n$ for the channel implemented by
    the protocol, \cref{lem:approx} now yields
    \begin{equation}
        \frac{1}{2} \norm{\widetilde{\cM}_n(\omega^{\otimes n})- \cM_n(\omega^{\otimes n})}_1 \leq \sqrt{2\delta_n+2\ee^{-2n^{1/3}}} = o(1)\, ,
    \end{equation}
    which proves the claim.    
\end{proof}

\begin{lemma}\label{lem:oracle}
    Given the list $(i_1,\dots i_K)$ and an index $s$, the conditional phase oracle
    \begin{equation}
        \ket{s}_C \ket{\Omega} \mapsto \ket{s}_C (I-2P^\perp_{i_s})\ket{\Omega}
    \end{equation}
    can be implemented by LOCC using $\log K + 2 \log d$ ebits. Here, $(\id-2P^\perp_{i_s})$ acts on the $i_s$ subsystem.
\end{lemma}
\begin{proof}
Without loss of generality assume that $d=\dim(B)\leq \dim(A)$ and that Alice has a quantum register $C$ indexing the public list. The two parties can then implement the controlled operation by first making a coherent copy of the basis index $s$ and then teleporting this register to Bob. Bob uses it to make a controlled swap between an initialized ancilla system and the selected subsystem $B_{i_s}$. Then he teleports this system to Alice. Alice now holds both halves of the selected subsystem and can apply the reflection. She then teleports the register back and Bob applies the same swap. In the end, Bob measures his index register in the Fourier basis. Alice afterwards corrects the resulting phase. This erases Bob's index register while preserving the coherence of Alice's register. 
Running this protocol consumes $\log K$ ebits to send the index and $2\log d$ for the teleportation to Alice and back.
\end{proof}

Before we present the actual algorithm performing the operation given in~\cref{eq_operation}, note that once we have a candidate position, we can use $2\log d$ ebits to teleport the system to Alice's side, perform a measurement in $\{P,\id - P\}$ and then teleport the system back. 
The verification determines whether a candidate position is marked, i.e.~whether $z_i=1$.
Whenever a previously undiscovered marked position is verified, we add it to the public set $S$. Subsequent lists are still sampled uniformly from all $n$ positions, but entries belonging to $S$ are treated as unmarked.
Thus each successful trial identifies a new marked position.

\begin{lemma}\label{lem:algo}
Let $0<p<\frac{1}{16}$ and $r_n=\lceil pn+n^{2/3}\rceil$. There exists a sequence of LOCC protocols using $nc(p)+o(n)$ ebits whose dilations have the form
\begin{equation}
        V_n=\sum_{z\in\{0,1\}^n} \Pi_z\otimes\ket{v_{n,z}}_{ZE} \, ,
\end{equation}  
where the output register $Z$ has an orthonormal basis $\{\ket{\hat z}_Z: \widehat{z} \in\{0,1\}^n\cup\{\bot\}\}$, $E$ contains a copy of the output label and the normalized states $\ket{v_{n,z}}$ satisfy
    \begin{equation}
        \delta_n\coloneqq
        \sup_{\substack{z\in\{0,1\}^n\\ |z|\leq r_n}} \left( 1- \norm{(\bra{z}_Z\otimes\id_E) \ket{v_{n,z}}_{ZE}}^2\right) = o(1) \, .
    \end{equation}   
\end{lemma}

\begin{proof}
\textbf{Description of the algorithm:}
Let $n$ be sufficiently large that $\frac{r_n}{n} < \frac{1}{16}$. We call a position $i\in\{1,\ldots,n\}$ \emph{marked} if $z_i=1$.
The algorithm searches for all marked positions and records those already found in a public set $S$. Since we do not know the actual number of marked positions, we start with an upper bound given by $j=r_n$. The justification for this choice is given in the correctness analysis below. 
Initially, we set $S=\varnothing$, and $j=r_n$ which are being updated during the algorithm. 
This is done in \emph{stages}. A stage consists of a fixed number of search attempts, called \emph{trials}, performed with the same estimate of $j$. Each trial applies Grover to a randomly sampled list of $K$ entries. This is explained in more detail below.
After running $N_j$ of these trials, we update our parameters as follows.

At the beginning of a stage with $j>0$, set
\begin{equation}
    j'=\left\lfloor \frac{999}{1000} j \right\rfloor, \qquad q=\frac{j}{n}, \qquad K=\left\lfloor\frac{16}{q}\right\rfloor, \qquad \textnormal{and} \qquad N_j=\left\lceil\frac{3}{2}(j-j') + \sqrt{j-j'}\ln n\right\rceil.
\end{equation}
Here $j'$ is the target upper bound for the next stage, $q$ is the current estimate of the remaining marked fraction, $K$ is the length of the list searched in each trial, and $N_j$ is the number of trials in the stage. The goal of the stage is to leave at most $j'$ marked positions outside $S$.

Each of the $N_j$ trials consists of the following steps. All classical random choices and verified discoveries are shared between the parties.

\begin{enumerate}
    \item Sample a fresh public list $(i_1,\ldots,i_K)$ of independent, uniformly distributed positions in $\{1,\ldots,n\}$, allowing repetitions. Positions already in $S$ are treated as unmarked.
    \item Set $m=\frac{\pi}{3\arcsin\sqrt q}$. To obtain an integer number of Grover iterations, let $M_-\leq m<M_+=M_-+2$ be the two consecutive odd integers surrounding $m$, and choose $M$ according to
    \begin{equation} \label{eq_M}
        \PP(M=M_-)=\frac{M_+-m}{2}
        \qquad \textnormal{and} \qquad 
        \PP(M=M_+)=\frac{m-M_-}{2}\, .
    \end{equation}
    Thus $M$ is odd and $\E[M]=m$.
    \item Alice prepares an index register $C$ in the uniform state
    \begin{equation}
        \ket{+}_C
        =\frac{1}{\sqrt K}\sum_{s=1}^K\ket{s}_C\, .
    \end{equation}
    The search then uses the phase oracle
    \begin{equation}
        \ket{s}_C
        \longmapsto
        (-1)^{z_{i_s}\mathbf{1}_{\{i_s\notin S\}}}\ket{s}_C\, ,
    \end{equation}
    which flips the phase precisely when the corresponding position is marked and has not yet been found. This oracle is implemented using \cref{lem:oracle}, with the reflection omitted for list entries already in $S$. Perform $(M-1)/2$ Grover iterations. Each iteration consists of this phase oracle followed by the local reflection $2\proj{+}_C-\id_C$.
    
    \item Measure $C$ in the computational basis. An outcome $s$ gives the candidate position $i_s$. If $i_s\notin S$, verify the candidate by performing the measurement $\{P,\id-P\}$ on that copy. Upon obtaining the outcome $\id-P$, add $i_s$ to $S$. In every other case, leave $S$ unchanged.
\end{enumerate}

    The parameters $j$, $j'$, $q$, $K$, and $N_j$ remain fixed throughout the stage, while $S$ is updated after each verified discovery. After the $N_j$ trials, replace $j$ by $j'$ and begin the next stage if $j>0$.
    When $j=0$, report the string $\widehat z$ defined by $\widehat z_i = 1$ for $i \in S$ and $0$ otherwise.
    To make the resource cost of the algorithm precise, fix an entanglement budget $b_n$ before starting the algorithm. Abort if the next oracle call or verification would make the total consumption exceed $b_n$.
    The cost analysis below will specify a budget $b_n=n c(p)+o(n)$ and show that the probability of such an abort vanishes.
    
    \paragraph{Correctness:}
    Let $k$ denote the actual fraction of marked positions outside $S$ and let $X$ denote the number of marked entries in a newly sampled list. $X$ also counts multiplicities and is therefore distributed as $\operatorname{Bin}(K,k)$. For given $M$ and $X$, the probability of finding a newly marked position is
    \begin{equation}
        P_{\mathrm{succ}} = \sin^2\left(M\arcsin\sqrt{\frac{X}{K}}\right).
    \end{equation}
    This follows from the analysis of Grover search with multiple marked entries~\cite{Boyer_1998}, applied to $X$ marked entries among $K$ entries after $(M-1)/2$ Grover iterations.
    
    In the ideal case $M=m$ and $X/K=q$, this probability is $\sin^2(\pi/3)=3/4$. As shown in \cref{lem:devi}, rounding and sampling reduce the success probability by less than $1/12$. 
    By \cref{lem:devi}, each trial finds a new marked position with probability at least $2/3$ whenever
       $ \frac{999}{1000} \, q<k\leq q  $.
    To bound the probability of failure of a whole stage, we get
    \begin{equation}
        \PP(\text{stage fails}) 
        \leq \PP\left( \operatorname{Bin}\Big(N_j,\frac{2}{3}\Big) <j-j' \right)\, .
    \end{equation}
    Here failure means that more than $j'$ marked positions remain outside $S$ at the end of the stage. By definition of $N_j$, we have
    \begin{equation}
        \frac{2}{3} N_j-(j-j')\geq\frac{2}{3}\sqrt{j-j'}\,\ln n \, ,
    \end{equation}
    which means that we can bound using Hoeffding's inequality.
    For sufficiently large $n$ we have
    $N_j\leq2(j-j')\ln n$. Hoeffding's inequality therefore yields
    \begin{equation}
        \PP (\text{stage fails}) \leq \exp\left( -\frac{2((2/3)N_j-(j-j'))^2}{N_j}\right) \leq n^{-\frac{4}{9}}.
    \end{equation}
    Starting from $|z|\leq r_n$, each successful stage establishes the upper bound used in the next stage. There are $O(\log n)$ stages, so a union bound gives
    \begin{equation}
        \PP(\text{some stage fails}) \leq O(\log n)\,n^{-\frac{4}{9}} = o(1)\, .
    \end{equation}
    If all stages succeed, no marked positions remain outside $S$ when $j=0$. Since verification prevents false discoveries, the reported string then satisfies $\widehat z=z$.
    This shows correctness uniformly over all $|z|\leq r_n$.

    We will now analyze the cost of our algorithm.
    At density $q=j/n$, unbiased rounding of $m$ gives
    \begin{equation}
        \mathbb E \left[\frac{M-1}{2} \right] = \frac{\pi}{6\arcsin\sqrt q}-\frac{1}{2} \leq\frac{\pi}{6\sqrt q}-\frac{1}{2}\,.
    \end{equation}
    Each oracle call costs $\log K+2\log d$ ebits, and verification
    costs at most $2\log d$ ebits. Since $K\leq16/q$, the expected
    cost of a trial is at most
    \begin{equation}
        f(q) \coloneqq \left(\frac{\pi}{3\sqrt q}+1\right)\log d + \left(\frac{\pi}{6\sqrt q}-\frac12\right) \log\frac{16}{q}.
    \end{equation}
    This function is positive and decreasing on the relevant interval.
    For a stage with $j'=\left \lfloor \frac{999}{1000} \, j\right \rfloor$,
    we thus obtain
    \begin{equation}
        (j-j') f(j/n) \leq n\int_{\frac{j'}{n}}^{\frac{j}{n}} f(t)\,\mathrm dt.
    \end{equation}
    
    Let $C_n$ denote the cost of all scheduled oracle calls together with one verification per trial.
    Since $N_j\leq\frac32(j-j')+\sqrt{j-j'}\,\ln n+1$, we get
    \begin{align}
        \E[C_n] &\leq \sum_{\text{stages }j}N_j f(j/n)\\
        &\leq \frac32\sum_{\text{stages }j}(j-j')f(j/n) + \sum_{\text{stages }j} \bigl(\sqrt{j-j'}\,\ln n+1\bigr)f(j/n)\\
        &\leq \frac{3n}{2}\int_0^{\frac{r_n}{n}}f(t)\,\mathrm dt+o(n) \,.
    \end{align}
    Here the stage intervals partition $(0,r_n/n]$.
    For the remainder, there are $O(\log n)$ stages, and $f(j/n)=O(\sqrt{n/j} \log n)$ and $\sqrt{j-j'}\leq\sqrt{j}$.
    The additional trials therefore contribute at most $O(\sqrt n(\log n)^3)=o(n)$. As $r_n/n\to p$, this gives
    \begin{equation}
        \E[C_n]\leq n c(p)+o(n)\, ,
    \end{equation}
    where
    \begin{align}
        c(p) =\frac{3}{2} \int_0^p f(t)\,\di t = \pi\sqrt p\log\frac{4\mathrm e d}{\sqrt p} +\frac{3}{2}p\log d-\frac{3}{4}p\log\frac{16\mathrm e}{p}.
    \end{align}
    
    It remains to enforce a fixed budget $b_n$. All stage lengths are predetermined, so we may independently sample all rounding choices before the execution of the algorithm. 
    
    In total, we have
    \begin{equation}
        \sum_{\text{stages }j} N_j = \frac{3}{2}r_n+O(\sqrt n\log n)=O(n) \, ,
    \end{equation}
    where we used $\sum_{\text{stages }j}\sqrt{j-j'}=O(\sqrt n)$, since the stage values $j$ decrease by a multiplicative factor.

    Changing one rounding choice changes the number of oracle calls by one, and hence changes $C_n$ by at most $\log(16n)+2\log d=O(\log n)$. Set the budget cutoff therefore to
     \begin{equation}
        b_n=\left\lceil \E[C_n]+n^{2/3}\log n \right\rceil \leq n c(p)+o(n)\,.
    \end{equation}
    Since the rounding choices are independent, Hoeffding's inequality then gives for some constant $\eta > 0$
    \begin{equation}
        \PP(C_n>b_n) \leq \ee^{-\eta n^{1/3}} = o(1)\, .
    \end{equation}
    Abort the protocol if the sampled schedule has $C_n>b_n$. The probability of this tends to zero. Together with the correctness analysis, this proves $\delta_n\to0$ uniformly over $|z|\leq r_n$.

    Within each subspace $\Pi_z$, the oracle calls leave the input state unchanged and verification has deterministic outcomes. The ancillary state therefore depends only on $z$, which gives the claimed dilation with an orthogonal record of the report in the
    environment.
\end{proof}

\begin{lemma}\label{lem:approx}
    Let $(\Pi_z)_{z\in\mathcal Z}$ be a finite family of orthogonal projectors on $\cH$ summing to the identity. Suppose a channel $\cN: \cH \mapsto \cH \otimes \cH_Z$ has an isometric dilation
    \begin{equation}
        V=\sum_{z\in\mathcal Z}\Pi_z\otimes\ket{v_z}_{ZE}\, .
    \end{equation}
    The register $Z$ stores a classical report $\widehat z\in\mathcal Z\cup\{\bot\}$, with $\bot$ denoting an abort, and $E$ contains a copy of this report.
    Let $e_z$ be the probability that $\widehat z\neq z$ in $\ket{v_z}_{ZE}$. Then for
    \begin{equation}
        \cM(X)=\sum_{z\in\mathcal Z}\Pi_zX\Pi_z \otimes\proj{z}_Z 
    \end{equation}
    and every state $\omega$ on $\mathcal H$, we have
    \begin{equation}
        \frac{1}{2} \norm{\cN(\omega)-\cM(\omega)}_1\leq\sqrt{2\sum_{z\in\mathcal Z}\tr[\Pi_z\omega]\,e_z}.
    \end{equation}
\end{lemma}

\begin{proof}
    For each $z$, write the component reporting the correct label as
    \begin{align}
        (\bra{z}_Z\otimes \id_E)\ket{v_z}_{ZE} =\sqrt{1-e_z}\,\ket{\xi_z}_E,
    \end{align}
    where $\ket{\xi_z}_E$ is normalized.
    If $e_z=1$, choose any unit vector in the environment subspace recording $z$, enlarging $E$ if necessary.
    Since the environment contains a copy of the report, the vectors $\ket{\xi_z}_E$ are mutually orthogonal.
    Consequently,
    \begin{align}
        W=\sum_{z\in\mathcal Z}
            \Pi_z\otimes\ket{z}_Z\ket{\xi_z}_E
    \end{align}
    is an isometric dilation of $\cM$.

    Let $\ket{\Omega}$ be a purification of $\omega$, and define
    \begin{align}
        \ket{\Psi}=(V\otimes \id_R)\ket{\Omega}
        \qquad \textnormal{and}\qquad
        \ket{\Phi}=(W\otimes \id_R)\ket{\Omega}\, ,
    \end{align}
    where $R$ is the purifying system.
    Orthogonality of the projectors gives
    \begin{align}
        \langle\Phi|\Psi\rangle
        =\sum_{z\in\mathcal Z} \tr[\Pi_z\omega]\sqrt{1-e_z}
        \geq\sum_{z\in\mathcal Z} \tr[\Pi_z\omega](1-e_z)
         =1-e \, ,
    \end{align}
    where we introduced $e = \sum_z \tr[\Pi_z\omega] e_z $.
    The trace distance between these pure states is then
    \begin{align}
        \frac{1}{2} \norm{\proj{\Psi} - \proj{\Phi} }_1
        =\sqrt{1-|\langle\Phi|\Psi\rangle|^2}
        \leq\sqrt{1-(1-e)^2}
        \leq\sqrt{2e}\, .
    \end{align}
    Tracing out $E$ and $R$ cannot increase the trace distance, proving the claim.
\end{proof}

\paragraph{Acknowledgements}
We acknowledge GPT 6 Astra for assistance in developing the proof.
D.S.~thanks Andreas Winter for discussions about the continuity of the distillable entanglement during the Banff workshop on additivity problems in July 2026.
L.S.~acknowledges support from the Quantum Center at ETH Zurich and SNSF Grant No. 200021E\_232425 and 2000-1E\_239000.
\appendix
\section{Technical results}
\begin{remark} \label{rmk_existence_unitary}
Let $\rho,\sigma \in \St(\cH)$ such that $\delta \coloneqq \frac{1}{2} \norm{\rho-\sigma}_1 < \lambda\coloneqq\lambda_{\min}^{+}(\rho)$.
Then there exist a unitary $U$ and a density matrix $\tau$ such that
\begin{align}
    \sigma=U\Big(\big(1-\frac{\delta}{\lambda}\big)\rho+\frac{\delta}{\lambda}\tau\Big)U^\dagger,
    \qquad \textnormal{and} \qquad 
    \norm{U-\id}_\infty\leq\frac{\delta}{\lambda-\delta} \, .
\end{align}
For $\delta>0$ and $b=\lambda-\delta>0$, and relative to $\cH=\supp\rho \oplus \ker\rho$ write the given states in block form as
\begin{align}
    \rho = \begin{pmatrix}R&0\\0&0\end{pmatrix},
    \qquad \textnormal{and} \qquad
    \sigma=\begin{pmatrix}B&C^\dagger\\C&D\end{pmatrix},
\end{align}
where $R$ is the restriction of $\rho$ to its support. Since $\norm{\rho-\sigma}_\infty\leq\delta$, $B\geq(1-\frac{\delta}{\lambda})R\geq b\id$ and $\norm C_\infty\leq\delta$. Define
\begin{align}
    X = \sqrt{B^2+C^\dagger C}\,,\qquad
    T = \begin{pmatrix}B&-bB^{-1}C^\dagger\\C & b\id\end{pmatrix},
    \qquad \textnormal{and} \qquad  A=B\oplus b\id \, .
\end{align}
Then
\begin{align} \label{eq_unitary_exits}
    T^\dagger T
    = X^2\oplus b^2(\id+CB^{-2}C^\dagger)\geq b^2\id,
    \qquad \textnormal{and} \qquad \norm{T-A}_\infty\leq\delta \, .
\end{align}
Thus the polar decomposition $T=U|T|$ has $U$ unitary and $|T|\geq b\id$. For any unit eigenvector $Ux=zx$, set $h=\langle x,|T|x\rangle\geq b$ and $\alpha=\langle x,Ax\rangle\geq b$. Then
\begin{align}
    \delta^2 \overset{\textnormal{\Cshref{eq_unitary_exits}}}{\geq} |\langle x,(T-A)x\rangle|^2 = |zh-\alpha|^2 = (h-\alpha)^2+h\alpha|z-1|^2 \geq b^2|z-1|^2 \, , \label{eq_unitary_exits_2}
\end{align}
where the final step uses that $h \alpha \geq b^2$. Hence, we find
\begin{align}
    \norm{U-\id}_\infty  = \max_{z \in \mathrm{spec}(U)} |z-1| \overset{\textnormal{\Cshref{eq_unitary_exits_2}}}{\leq} \frac{\delta}{b} \, .
\end{align}
The first block column of $U$ is $\binom BC X^{-1}$.
By the Schur complement,
\begin{equation}
    \sigma \geq \binom BC B^{-1}\begin{pmatrix}B&C^\dagger\end{pmatrix} = U(XB^{-1}X\oplus0)U^\dagger.
\end{equation}
Operator monotonicity of the square root~\cite[Theorem V.1.9]{bhatia_book} gives $X\geq B$ and thus
\begin{equation}
    XB^{-1}X-B=2(X-B)+(X-B)B^{-1}(X-B)\geq 0 \, .
\end{equation}
Consequently $\sigma\geq(1 - \frac{\delta}{\lambda})U\rho U^\dagger$, so $\tau=[U^\dagger\sigma U-(1-\frac{\delta}{\lambda})\rho]/\frac{\delta}{\lambda}$ is positive with trace one.
\end{remark}

\begin{lemma}\label{lem:devi}
Using the notation of~\cref{lem:algo}, we have
\begin{equation}
    \E_{M,X}\Big[ \,\sin^2\left(M\arcsin\sqrt{X/K}\right) \Big] > \frac{2}{3}
    \qquad \text{whenever} \qquad \frac{999}{1000} \,q<k\leq q.
\end{equation}
\end{lemma}
\begin{proof}
Let us define $f_M(t)=\sin^2(M\arcsin\sqrt t)$. Since the ideal case has $f_m(q) = \sin^2(\pi/3)=3/4$, we do a Taylor expansion around this value in $m$. 
For this, we need to control the expectation value between the rounded $M$ and $m$. We find
\begin{equation}\label{eq:rounding}
\E_M[(M-m)] = 0\, , \quad 
\E_M[(M-m)^2]\leq 1\,, \quad 
\E_M|M-m|^3\leq 1\,,  \quad \textnormal{and} \quad
\E_M[(M-m)^4]\leq 4 \, ,
\end{equation}
which can be seen by using that two consecutive odd integers are two units apart. The Taylor expansion then gives
\begin{equation}\label{eq:phase}
    \E_M[f_M(q)] 
    \geq\frac{3}{4}-\frac{1}{2} \arcsin^2\sqrt q -\frac{2}{3}\arcsin^3\sqrt q 
    \geq\frac{3}{4}-\frac{1}{30} - \frac{2}{45\sqrt{15}} 
    >\frac{3}{4} - \frac{9}{200}\, ,
\end{equation}
where the last step used $\arcsin\sqrt{q}\leq 1/\sqrt{15}$. To bound the error introduced by sampling from a list, we use the bounds
\begin{equation}\label{eq:deriv}
    \norm{f_M'}_\infty\leq M^2, \qquad \norm{ f_M''}_\infty \leq\frac{2}{3} M^2(M^2-1) \, .
\end{equation}
These follow from Markov's inequality for polynomial derivatives~\cite[Section~1.1]{Shadrin2004} applied on $[0,1]$. Taylor's theorem around $k$ with the mean and the variance of $X/K$ then gives
\begin{equation}
    \E_X[f_M(X/K)] \geq f_M(k) -\frac{M^2(M^2-1)}{3}\frac{k(1-k)}{K} \,.
\end{equation}
Expanding the polynomial in $M-m$ and using \cref{eq:rounding}, we get
\begin{equation}
    \E_M[M^2(M^2-1)] \leq m^4+5m^2+4m+3 \leq m^4+\frac{25}{4}m^2,
\end{equation}
where the last inequality holds for $m\geq4$. Since $k(1-k)\leq q$, $K\geq(16-q)/q$, and $m^2q\leq\pi^2/9$, it follows that
\begin{equation}\label{eq:sampling}
\E_{M,X}[f_M(X/K)] \geq\E_M[ f_M(k)] -\frac{\frac{\pi^4}{81}+\frac{25}{4}\frac{\pi^2}{9}q}{3(16-q)}
    \geq\E_M[f_M(k)]-\frac{7}{200} \, .
\end{equation}
For the second inequality, we used $q\leq 1/16$ and $\pi^2<10$.

Now we relate the actual fraction $k$ to our guessed one $q$. If $\frac{999}{1000} \,q<k\leq q$, the first bound in \cref{eq:deriv} and $\E[M^2]\leq m^2+1$ imply
\begin{equation}
    \E_M[f_M(k)]
    \!\geq\!\E_M[f_M(q)]-(q-k)\E_M[M^2]
    \!\geq\!\E_M[f_M(q)] -\left(1-\frac{999}{1000} \right)\left(\frac{\pi^2}{9}+q\right)
    \!\geq\!\E_M[f_M(q)]-\frac6{5000}\, ,
\end{equation}
using $\pi^2/9+q<6/5$. Combining this with~\cref{eq:phase} and~\cref{eq:sampling} gives
\begin{equation}
        \E_{M,X}[f_M(X/K)] \geq\frac34-\frac{9}{200}-\frac{7}{200}-\frac{6}{5000} >\frac{2}{3} \,,
\end{equation}
which proves the lemma.
\end{proof}

\begin{lemma}\label{lem:bhatia}
Let $\omega \in \St(\cH)$, $0\leq A\leq\omega$, $p=1-\tr[A]$, and let $C=\mathbb C^2$ with standard basis $|0\rangle_C,|1\rangle_C$. There exists an orthogonal projector $P$ on $\mathcal H\otimes C$ such that
\begin{align}
    A = (\id\otimes\bra{0}_C)P(\omega\otimes\proj{0}_C) P(\id\otimes\ket{0}_C) \qquad \textnormal{and} \qquad 
    p \geq \tr\bigl[(\id-P)(\omega\otimes\proj{0}_C)\bigr] \, .
\end{align}
\end{lemma}
\begin{proof} 
Consider
\begin{align} \label{eq_choice_K}
 K=\omega^{-1/2}(\omega^{1/2} A\omega^{1/2})^{1/2}\omega^{-1/2}\, ,
\end{align}
which satisfies $0\leq K\leq \id$. This follows from the operator monotonicity of the square root~\cite[Theorem V.1.9]{bhatia_book} together with $\omega^{1/2}A\omega^{1/2}\le\omega^2$. 
The identity $K\omega K=A$ follows directly from~\cref{eq_choice_K}.
Set
\begin{align}
 P=\begin{pmatrix}K&S\\S&\id-K\end{pmatrix} \qquad \textnormal{for} \qquad  S=\sqrt{K(\id-K)} \, .
 \end{align}
The identities $KS=SK$ and $S^2=K-K^2$ imply $P^2=P=P^\dagger$.
We then observe
\begin{align}
 &\hspace{-20mm}(\id \otimes|0\rangle\langle0|_C)P(\omega\otimes \proj{0}_C)P(\id \otimes \proj{0}_C) \nonumber \\
 &=(\id \otimes \proj{0}_C )P(\id \otimes \proj{0}_C)(\omega\otimes\proj{0}_C)(\id \otimes \proj{0}_C)P(\id \otimes \proj{0}_C) \\
 &=(K \otimes \proj{0}_C)(\omega\otimes \proj{0}_C)(K \otimes \proj{0}_C) \\
 &=(K\omega K)\otimes \proj{0}_C \\
 &=A\otimes \proj{0}_C \, .
\end{align}
Since $0\leq K \leq \id$ we have $K\geq K^2$ and hence $\tr[\omega K^2]  \leq \tr[K \omega]$.
As a result, 
\begin{align}
 \tr[(\id-P)(\omega\otimes \proj{0}_C)] = 1 - \tr[\omega K] 
 \leq 1-\tr[\omega K^2] = 1-\tr[A] = p \, . \label{eq_final_proof_eq}
\end{align}
\end{proof}

\begin{lemma} \label{lem_scalar}
Let $d \in \N$, $x \in [0,\frac{1}{16}]$, and $c(\cdot)$ be the function defined in~\cref{eq_c_function}. Then
\begin{align} \label{eq_scalar_bound}
    \min\Big \{ \log d, x \log d + (1+x) c\Big(\frac{x}{1+x}\Big) \Big \}
    \leq 4 \sqrt{x} \log d + 4 h(\sqrt{x}) \, .
\end{align}
\end{lemma}
\begin{proof}
The endpoints $x=0$ and $x=\frac{1}{16}$ are immediate, so assume
$0<x<\frac{1}{16}$. If
\begin{align}
 \log d\leq\frac{4h(\sqrt x)}{1-4\sqrt x} \, ,
\end{align}
the bound follows from the first term in the minimum. 
Otherwise, subtract the desired right-hand side from the second term in the minimum. The result is affine and decreasing in $\log d$, since
\begin{align}
 \pi\sqrt{1+x}+\frac{5}{2}\sqrt x
 \leq\frac{22}{7}\frac{33}{32}+\frac58
 =\frac{433}{112}<4 \, .
\end{align}
It therefore suffices to consider $\log d=\frac{4h(\sqrt x)}{1-4\sqrt x}$ which simplifies~\cref{eq_scalar_bound} to 
\begin{align}\label{eq:alternative-scalar-goal}
 4\Big(1\!-\!\pi\sqrt{x(1\!+\!x)}\!-\!\frac{5}{2}x\Big) \frac{\ln2}{\sqrt x}h(\sqrt x) \!-\!(1\!-\!4\sqrt x)\left(\Big(\pi\sqrt{1\!+\!x}\!-\!\frac{3}{2}\sqrt x\Big)
       \ln\frac{4\mathrm \ee\sqrt{1\!+\!x}}{\sqrt x} \!+\!\frac{3}{4}\sqrt x\right)\geq 0 \, ,
\end{align}
for $x \in (0,\frac{1}{16})$. This can be verified with standard basic analysis tools (e.g.~via tedious series expansions).
\end{proof}


\bibliographystyle{arxiv_no_month}
\bibliography{bibliofile}

\begin{thebibliography}{10}

\bibitem{AaronsonAmbainis2005}
S.~Aaronson and A.~Ambainis.
\newblock Quantum search of spatial regions.
\newblock {\em Theory of Computing}, 1(4):47--79, 2005.
\newblock
  \texttt{\href{http://dx.doi.org/10.4086/toc.2005.v001a004}{DOI:\,10.4086/toc.2005.v001a004}}.

\bibitem{hitch_book}
C.~D. Aliprantis and K.~C. Border.
\newblock {\em Infinite Dimensional Analysis}.
\newblock Springer, 3 edition, 2006.
\newblock
  \texttt{\href{http://dx.doi.org/10.1007/3-540-29587-9}{DOI:\,10.1007/3-540-29587-9}}.

\bibitem{bennett96}
C.~H. Bennett, H.~J. Bernstein, S.~Popescu, and B.~Schumacher.
\newblock Concentrating partial entanglement by local operations.
\newblock {\em Phys. Rev. A}, 53:2046--2052, 1996.
\newblock
  \texttt{\href{http://dx.doi.org/10.1103/PhysRevA.53.2046}{DOI:\,10.1103/PhysRevA.53.2046}}.

\bibitem{BBPSSW96}
C.~H. Bennett, G.~Brassard, S.~Popescu, B.~Schumacher, J.~A. Smolin, and W.~K.
  Wootters.
\newblock Purification of noisy entanglement and faithful teleportation via
  noisy channels.
\newblock {\em Phys. Rev. Lett.}, 76:722--725, 1996.
\newblock
  \texttt{\href{http://dx.doi.org/10.1103/PhysRevLett.76.722}{DOI:\,10.1103/PhysRevLett.76.722}}.

\bibitem{bennett_mixed-state_1996}
C.~H. Bennett, D.~P. {DiVincenzo}, J.~A. Smolin, and W.~K. Wootters.
\newblock Mixed-state entanglement and quantum error correction.
\newblock {\em Physical Review A}, 54(5):3824--3851, 1996.
\newblock
  \texttt{\href{http://dx.doi.org/10.1103/PhysRevA.54.3824}{DOI:\,10.1103/PhysRevA.54.3824}}.

\bibitem{bhatia_book}
R.~Bhatia.
\newblock {\em Matrix Analysis}.
\newblock Springer, 1997.
\newblock
  \texttt{\href{http://dx.doi.org/10.1007/978-1-4612-0653-8}{DOI:\,10.1007/978-1-4612-0653-8}}.

\bibitem{Boyer_1998}
M.~Boyer, G.~Brassard, P.~Høyer, and A.~Tapp.
\newblock Tight bounds on quantum searching.
\newblock {\em Fortschritte der Physik}, 46(4-5):493–505, 1998.
\newblock
  \texttt{\href{http://dx.doi.org/10.1002/(sici)1521-3978(199806)46:4/5<493::aid-prop493>3.0.co;2-p}{DOI:\,10.1002/(sici)1521-3978(199806)46:4/5<493::aid-prop493>3.0.co;2-p}}.

\bibitem{LOCC14}
E.~Chitambar, D.~Leung, L.~Man{\v c}inska, M.~Ozols, and A.~Winter.
\newblock Everything you always wanted to know about {LOCC} (but were afraid to
  ask).
\newblock {\em Communications in Mathematical Physics}, 328(1):303--326, 2014.
\newblock
  \texttt{\href{http://dx.doi.org/10.1007/s00220-014-1953-9}{DOI:\,10.1007/s00220-014-1953-9}}.

\bibitem{DW05}
I.~Devetak and A.~Winter.
\newblock Distillation of secret key and entanglement from quantum states.
\newblock {\em Proceedings: Mathematical, Physical and Engineering Sciences},
  461(2053):207--235, 2005.
\newblock Available online: \url{http://www.jstor.org/stable/30046924}.

\bibitem{fekete23}
M.~Fekete.
\newblock {\"U}ber die {V}erteilung der {W}urzeln bei gewissen algebraischen
  {G}leichungen mit ganzzahligen {K}oeffizienten.
\newblock {\em Mathematische Zeitschrift}, 17(1):228--249, 1923.

\bibitem{winter05}
B.~Groisman, S.~Popescu, and A.~Winter.
\newblock Quantum, classical, and total amount of correlations in a quantum
  state.
\newblock {\em Phys. Rev. A}, 72:032317, 2005.
\newblock
  \texttt{\href{http://dx.doi.org/10.1103/PhysRevA.72.032317}{DOI:\,10.1103/PhysRevA.72.032317}}.

\bibitem{hastings09}
M.~B. Hastings.
\newblock Superadditivity of communication capacity using entangled inputs.
\newblock {\em Nature Physics}, 5(4):255--257, 2009.
\newblock
  \texttt{\href{http://dx.doi.org/10.1038/nphys1224}{DOI:\,10.1038/nphys1224}}.

\bibitem{HHT01}
P.~M. Hayden, M.~Horodecki, and B.~M. Terhal.
\newblock The asymptotic entanglement cost of preparing a quantum state.
\newblock {\em Journal of Physics A: Mathematical and General}, 34(35):6891,
  2001.
\newblock
  \texttt{\href{http://dx.doi.org/10.1088/0305-4470/34/35/314}{DOI:\,10.1088/0305-4470/34/35/314}}.

\bibitem{Horodecki-Horodecki}
M.~Horodecki and P.~Horodecki.
\newblock Reduction criterion of separability and limits for a class of
  distillation protocols.
\newblock {\em Phys. Rev. A}, 59:4206--4216, 1999.
\newblock
  \texttt{\href{http://dx.doi.org/10.1103/PhysRevA.59.4206}{DOI:\,10.1103/PhysRevA.59.4206}}.

\bibitem{SS26Tele}
L.~Schmitt and D.~Sutter.
\newblock Essentially optimal gate teleportation, 2026.
\newblock Available online: \url{https://arxiv.org/abs/2608.24345}.

\bibitem{Shadrin2004}
A.~Shadrin.
\newblock Twelve proofs of the {Markov} inequality.
\newblock In D.~K. Dimitrov, G.~Nikolov, and R.~Uluchev, editors, {\em
  Approximation Theory: A Volume Dedicated to {Borislav Bojanov}}, pages
  233--298. Marin Drinov Academic Publishing House, Sofia, 2004.
\newblock Available online:
  \url{https://www.damtp.cam.ac.uk/user/na/people/Alexei/papers/markov.pdf}.

\bibitem{shor_additivity_04}
P.~W. Shor.
\newblock Equivalence of additivity questions in quantum information theory.
\newblock {\em Communications in Mathematical Physics}, 246(3):453--472, 2004.
\newblock
  \texttt{\href{http://dx.doi.org/10.1007/s00220-003-0981-7}{DOI:\,10.1007/s00220-003-0981-7}}.

\bibitem{Vidal02}
G.~Vidal.
\newblock On the continuity of asymptotic measures of entanglement, 2002.
\newblock Available online: \url{https://arxiv.org/abs/quant-ph/0203107}.

\bibitem{wilde_book}
M.~M. Wilde.
\newblock {\em Quantum Information Theory}.
\newblock Cambridge University Press, 2013.
\newblock
  \texttt{\href{http://dx.doi.org/10.1017/9781316809976}{DOI:\,10.1017/9781316809976}}.

\bibitem{win16}
A.~Winter.
\newblock Tight uniform continuity bounds for quantum entropies: Conditional
  entropy, relative entropy distance and energy constraints.
\newblock {\em Communications in Mathematical Physics}, 347(1):291--313, 2016.
\newblock
  \texttt{\href{http://dx.doi.org/10.1007/s00220-016-2609-8}{DOI:\,10.1007/s00220-016-2609-8}}.

\end{thebibliography}

\end{document}